\documentclass[pre,superscriptaddress,showpacs,twocolumn]{revtex4}
\usepackage{graphicx}
\usepackage{amssymb}
\usepackage{epstopdf}

\begin{document}
\title{Optimization by Quantum Annealing: Lessons from hard 3-SAT cases}

\author{Demian A. Battaglia}
\email{battagli@sissa.it}
\affiliation{International School for Advanced Studies (SISSA), and INFM
Democritos National Simulation Center, Via Beirut 2-4, I-34014 Trieste, Italy}

\author{Giuseppe E. Santoro}
\email{santoro@sissa.it}
\affiliation{International School for Advanced Studies (SISSA), and INFM
Democritos National Simulation Center, Via Beirut 2-4, I-34014 Trieste, Italy}
\affiliation{International Center for Theoretical Physics (ICTP), Trieste, Italy}

\author{Erio Tosatti}
\email{tosatti@sissa.it}
\affiliation{International School for Advanced Studies (SISSA), and INFM
Democritos National Simulation Center, Via Beirut 2-4, I-34014 Trieste, Italy}
\affiliation{International Center for Theoretical Physics (ICTP), Trieste, Italy}

\begin{abstract}
The Path Integral Monte Carlo simulated Quantum Annealing algorithm is applied to the 
optimization of a large hard instance of the Random 3-SAT Problem ($N=10000$). 
The dynamical behavior of the quantum and the classical annealing are compared,
showing important qualitative differences in the way of exploring the complex 
energy landscape of the combinatorial optimization problem.
%The ``genetic-like'' nature of the Path Integral Monte Carlo scheme is as well clarified, 
%in order to distinguish between genuinely quantum and purely classical effects, due to the
%specific simulation technique.
At variance with the results obtained for the Ising spin glass
and for the Traveling Salesman Problem, in the present case the linear-schedule 
Quantum Annealing performance is definitely worse than Classical Annealing. 
Nevertheless, a quantum cooling protocol based on field-cycling and able to outperform 
standard classical simulated annealing over short time scales is introduced.
\end{abstract}

\maketitle

%++++++++++++++++++++++++++++++++++++++++++++++++++++++++++++++++++++++++++++++++++++++++
\section{Introduction}
%++++++++++++++++++++++++++++++++++++++++++++++++++++++++++++++++++++++++++++++++++++++++

The aim of theoretical Quantum Computation is to understand 
how the peculiar properties of quantum states can be exploited 
in order to encode and manipulate information in a way which is
quantitatively superior to classical computers \cite{QC}.  
The elementary unit of quantum information, the so-called
\textit{qubit}, is in general assumed to be a normalized 
vector in a Hilbert space $\mathcal{H}$ spanned by two orthogonal 
eigenstates, conventionally labeled $\left|0\right>$ and
$\left|1\right>$, and it is then completely determined by giving 
two real numbers (a component and a relative phase). 
In principle, a quantum computation can be seen as a sequence of appropriate 
unitary transformations operating on sets of qubits, which requires having
the quantum state of the system under control for a sufficiently long time.

A completely different point of view is shared by a group of quantum 
computation paradigms that could be collectively referred to as
Quantum Ground State Search techniques: two examples 
are given by the Quantum Adiabatic Evolution technique and its variants 
\cite{Adiabatic1,Adiabatic2,Adiabatic_numberP,QA_chem,QA_Kadowaki} and by the semi-classical 
simulated Quantum Annealing algorithms \cite{QA_ising,QA_ising_long,QA_tsp}. 
The basic idea common to all these approaches is to simulate 
the externally driven evolution of an artificial quantum system, whose 
ground states encode in some way the solutions of specific computational problems. 
The underlying hypothesis is that a quantum system can dynamically explore the state 
space more efficiently than its classical counterpart. The positive role played by
quantum fluctuations has been verified in a number of different cases, both in real 
experiments \cite{QA_aeppli} and in numerical simulations \cite{QA_ising,QA_tsp}. 
The Path-Integral Monte Carlo Quantum Annealing (PIMC-QA) algorithm has been 
applied with success, for instance, to the optimization of instances 
of the Ising spin glass \cite{QA_ising} and of the Traveling Salesman Problem \cite{QA_tsp}. 
Here the quantum algorithm outperformed classical Simulated 
Annealing \cite{CA_original} and the convergence time was
significantly shorter, even after taking in account the extra cost of 
simulating a quantum computation on a classical computer \cite{QA_ising,QA_tsp}.

Nevertheless, very few results of universal validity are known to date,
and theoretical indications are poor and sometimes contradictory
\cite{Adiabatic2,Vazirani,Colin,tls_stella}.
In particular, it has not been proved that a good quantum 
optimization pathway must always exist in general. In the cases in which 
a quantum procedure proved to be more efficient than the best classical 
one, the superior quantum performance cannot be explained quantitatively. 

In this paper, we will present a careful analysis of the PIMC-QA 
applied to a hard instance of the random MAX-3-SAT problem \cite{Garey_Johnson}.
The main emphasis will not be on performance, but rather 
on understanding the properties that make simulated Quantum 
Annealing so different from its classical counterpart. 
Because we are here mainly interested in the comparison between algorithmic aspects, 
potentially relevant physical issues as the study of aging and 
memory/rejuvenation phenomena \cite{cugliandolo,bouchaud} will not be addressed directly.

The hardness of the chosen 3-SAT instance deserves some comments. 
It is common, in the quantum computation literature,
to analyze the performance of newly introduced algorithms by testing
them on simplified toy-problems of reduced size; this is essentially due to 
the demanding memory and time required by the 
classical simulation of a quantum computer. We decided instead to
attempt the optimization of a true ``competition problem'',
of a kind known to threaten seriously even the best heuristics known for the 3-SAT
problem, namely WalkSAT \cite{Walksat}. 
Here smaller size problems might indeed still be difficult,
but the dynamics would be dominated by large finite-size-fluctuation effects \cite{SP-Y}
and misleading indications could be obtained about the general properties of the quantum
optimization of a glassy energy landscape, which becomes really ``hard'' only in the 
thermodynamical limit.

For a hard 3-SAT problem, one may wonder now if the quantum system is actually able to tunnel 
across the extensive walls encircling the deep and scattered 
ground state valleys. Here we expect that both classical and quantum
annealing will be trapped by local minima lying well above 
the glassy threshold levels \cite{Montanari,SP-Y}, preventing to access the very 
low-energy region where clustering of states takes place.
The comparison between the simulated classical and quantum relaxations will
nevertheless show that the two algorithms explore sectors of the phase space 
characterized by different geometrical properties. 
As we shall see, the quantum pathways tend 
to visit basins of attraction with a considerably larger number of flat directions and 
this feature will turn out to be highly counterproductive 
for an efficient optimization in the specific case of the 3-SAT landscape. 
The poor performance of quantum search techniques applied to the
3-SAT problem, envisaged already in Ref.~\cite{Colin}, will then find here a striking
confirmation.

The paper is organized as follows. Sects.~\ref{3sat:sec} and \ref{pimc:sec} will 
give respectively a brief description of the combinatorial 
optimization problem under study and of the employed algorithms.  
In Sec.~\ref{num_exp:sec} a first set of numerical experiments will be presented, 
while the analogy between PIMC-QA and genetic-like algorithms will be pointed out in
Sec.~\ref{MC_dynamics:sec}. 
In Sec.~\ref{autocorr:sec} the differences between quantum and classical Monte Carlo
dynamics will be highlighted by means of an autocorrelation analysis, and of a 
characterization of the landscape geometry.
In Sec.~\ref{field_cycling:sec} a more sophisticated cooling schedule will be introduced, 
allowing the quantum algorithm to outperform classical Simulated Annealing, 
at least over short simulation times. 
In Sec.~\ref{conclusions:sec}, finally, concluding comments about open problems and possible 
future lines of development will be presented.

%++++++++++++++++++++++++++++++++++++++++++++++++++++++++++++++++++++++++++++++++++++++++
\section{The Random $3$-SAT Problem}
\label{3sat:sec}
%++++++++++++++++++++++++++++++++++++++++++++++++++++++++++++++++++++++++++++++++++++++++

Under many aspects, the 3-SAT problem can be considered as the prototype 
of most of the hard combinatorial optimization problems: 3-SAT was indeed the first 
problem for which a direct proof of \textbf{NP}-completeness was ever obtained. 
Since then, the easiest way of proving the \textbf{NP}-completeness of some new 
problem has been to show its equivalence with some 3-SAT instance \cite{Garey_Johnson}.
Furthermore, 3-SAT is also of considerable practical relevance in many
engineering applications, notably in the field of Artificial
Intelligence and Automated Planning \cite{planning1,planning2,ai_1,ai_2}. 

In order to state the problem, consider a set of $N$ boolean variables
$z_1,\cdots,z_N$, where $z_i=1 \; \mbox{or}\; 0$ ('True' or 'False').
Denoting by $\zeta_i$ the variable $z_i$ or its negation $\bar{z_i}$,
one then considers the disjunction (logical OR) of 3 variables
${C}=(\zeta_i \vee \zeta_j \vee \zeta_k)$, which is called a $3$-\,{\em clause}.
The random 3-SAT problem consists in deciding if the conjunction (logical AND) 
of $M$ different clauses ${C}_1 \wedge {C}_2 \cdots \wedge {C}_M$ 
-- each clause being formed by 3 variables extracted at random among the $N$ available, and
appearing negated or directed with uniform probability --
can be simultaneously satisfied by a truth value assignment $\{z_i\}$.
The optimization version of 3-SAT (the so called MAX-3-SAT problem) is obtained if 
one is interested only in finding a truth value assignment which minimizes the
number of violated clauses.
If we associate an Ising spin variable $S_i=(-1)^{z_i}$ to each Boolean variable $z_i$, 
we can assign to any clause $C_a$ involving three variables $z_i, z_j, z_k$
an energy $E_a$ given by
\begin{equation}\label{SAT_hamiltonian_term}
E_a = \frac{\left(1+J_{a,i}S_{i}\right)\left(1+J_{a,j}S_{j}\right)\left(1+J_{a,k}S_{k}\right)}{8}
\;,
\end{equation}
where the coupling $J_{a,i}$ assumes the value -1 if the variable $z_i$
appears negated in clause $a$, +1 otherwise.
Evidently $E_a=0$ if the corresponding clause is satisfied, $E_a=1$ otherwise.
Therefore, each given realization of the $M$ random clauses is associated 
to the Hamiltonian of a spin system with quenched disorder:
\begin{equation}\label{SAT_ham}
\mathcal{H}\left(\{S_i\}\right) = \sum_{a=1}^M E_a \;,
\end{equation}
which simply counts the number of violated clauses.
Notice that the same variable appears typically in more than one clause, sometimes negated,
sometimes not, giving thus origin potentially to frustration.

Statistical mechanics techniques can be used to determine the \textit{phase diagram} 
of the Random 3-SAT problem \cite{TCS,SP_science,SP_long}.
The main parameter in this phase diagram (determining the hardness of a formula) is the 
ratio $\alpha=M/N$ between the number, $M$, of clauses and the number, $N$, of variables.
The probability that a satisfying assignment exists becomes one in the thermodynamical 
limit for $\alpha<\alpha_c\simeq 4.267$, but vanishes for larger $\alpha$'s. 
Furthermore, for $\alpha_c>\alpha>\alpha_G \simeq 4.15$, the ground-state configurations 
group in many clusters well separated among each other 
(one-step replica symmetry breaking \cite{PMV}), hidden to any local search 
heuristic by an exponentially larger number of metastable states (\textit{threshold states}).  
A lower bound to the energy of the threshold states acting as entropic traps 
is provided by the so-called Gardner energy \cite{Montanari}.

Numerical experiments \cite{SAT_phase_trans_1,SAT_phase_trans_2} are in agreement with the 
analytical predictions \cite{TCS,SP_science,SP_long}, and confirm that instances of 
random 3-SAT sampled in the vicinity of the SAT/UNSAT boundary $\alpha_c$ are typically hard.
The trapping effect induced by the threshold states cannot be neglected when 
the instance-size is large ($N\geq 10000$) and large statistical fluctuations become 
sufficiently rare \cite{SP-Y}. 
Smaller random formulas are, on the other hand, often easily solvable by classical 
simulated annealing. 
As already pointed out in the introduction, the efficiency of any new heuristics can 
then be significantly tested only over 3-SAT samples of a rather large size, which 
is considerably demanding for long quantum simulation times. 

%++++++++++++++++++++++++++++++++++++++++++++++++++++++++++++++++++++++++++++++++++++++++
\section{The Path-Integral Monte Carlo scheme}
\label{pimc:sec}
%++++++++++++++++++++++++++++++++++++++++++++++++++++++++++++++++++++++++++++++++++++++++

Let us consider the classical spin model described by the Hamiltonian (\ref{SAT_ham}).
In the well known classical Simulated Annealing scheme \cite{CA_original}, the configuration of
the system is randomly initialized at an initial temperature $T_0$ and
it is then updated according to, for instance, a standard Metropolis Monte Carlo rule:
\begin{equation}\label{acceptance}
P_t(\mbox{flip})=\min\left[1, \exp\left(-\frac{\Delta E}{T_t}\right)\right],
\end{equation}  
where $P_t(\mbox{flip})$ is the probability of accepting a given spin flip at time $t$, 
$\Delta E$ is the energy change induced by the flip and $T_t$ is the temperature at time $t$. 
The temperature is gradually reduced according to a specific schedule that can be rather
influent on the performance of the algorithm (see, for instance, Refs.\ \cite{aarts,ictp_CA}). 
In the present paper we shall concentrate our attention on simple linear schedules, where 
the temperature is reduced by a fixed amount $\Delta T$ every $\delta$ Monte Carlo 
complete sweeps of the systems 
(we chose $\delta=2$ in our experiments, while $\Delta T$ was completely determined by 
the initial value $T_0$ and by the maximum number of iterations). 
This is not in general the best performing choice, but its
simplicity makes the comparison with quantum annealing very transparent. 

The classical Hamiltonian $\mathcal{H}\left(\{S_i\}\right)$, in analogy with the Ising
glass case \cite{QA_ising,QA_ising_long}, can be turned into a quantum Hamiltonian, 
by substituting each classical
Ising spin $S_i$ with the third component of a Pauli SU(2) spin operator
$\hat{\sigma}^z_i$, and by introducing a perturbing transverse field $\Gamma$:
\begin{equation}\label{quantum_ham}
\mathcal{H}_{\mbox{\tiny Quantum\normalsize}}=
\mathcal{H}\left(\left\{\hat{\sigma}_i^z\right\}\right) - \Gamma\sum_i\hat{\sigma}^x_i
\end{equation}
The $\Gamma$-dependent term plays the role of ``kinetic energy'', inducing quantum fluctuations
in the spin orientation. 
The basic idea of Quantum Annealing is to drive the system (\ref{quantum_ham})
toward its ground state by adiabatically varying the intensity of the 
perturbation field between two extremal values $\Gamma_0$ and $\Gamma_f$, -- 
instead of gradually reducing the temperature, like in the case of thermal Simulated Annealing. 

In the Path Integral Monte Carlo (PIMC) scheme, the statistical-mechanical behavior 
of the quantum spin model (\ref{quantum_ham}) is approximately turned into a classical
simulation problem
by resorting to the Suzuki-Trotter transformation \cite{suzuki_trotter,quantum_and_RSB}. 
Let us consider the following classical model with $PN$ degrees of freedom:
\begin{equation}\label{suzuki_ham}
\mathcal{H}_{\mbox{\tiny ST\normalsize}}=
  \frac{1}{P} \sum_{\rho=1}^P\mathcal{H}\left(\left\{S_{i, \rho}\right\}\right) - 
  J_\Gamma\sum_{\rho=1}^{P}\sum_i S_{i, \rho}S_{i,\rho+1}
\end{equation}
The system (\ref{suzuki_ham}) can actually be seen as composed of $P$ replicas (Trotter replicas)  
$\left\{S_{i,\rho},\rho=1\cdots P\right\}$ of the original classical configuration 
$\left\{S_i\right\}$ at an effective quantum temperature $T_q = PT$, coupled among them 
by a nearest-neighbor transverse ferromagnetic coupling $J_\Gamma$. 
Periodic boundary conditions must be imposed along the transverse (Trotter) direction.
The intensity of $J_\Gamma$ is related to the strength of the quantum perturbation term 
and to the temperature $T = T_q/P$ by the following relation:
\begin{equation}\label{transverse_coupling}
J_\Gamma=-\frac{T}{2}\ln\tanh{\left(\frac{\Gamma}{PT}\right)} > 0 \;.
\end{equation}
When $P$ goes to infinity, the partition functions of the Hamiltonians (\ref{quantum_ham}) 
and (\ref{suzuki_ham}) become identical, and the statistical-mechanical properties of the two
systems become perfectly equivalent.
One can then hope to simulate the relaxation dynamics of (\ref{quantum_ham}) 
by applying a classical Metropolis algorithm to a Suzuki-Trotter transformed 
Hamiltonian in Eq.~\ref{suzuki_ham} \cite{QA_ising}. 
The accuracy of the quantum simulation will increase 
for larger values of $P$, but the memory and time requirements will also increase accordingly.

In the simplest possible Monte Carlo implementation, one tries to flip independently each 
of the $PN$ spins of the equivalent classical system, but it is common \cite{kikuchi} 
to perform in addition also \textit{global moves} in which the spins $S_{i,\rho}$ 
corresponding to a given site $i$ are simultaneously flipped in all the Trotter replicas
($\rho=1\cdots P$). 
This type of move does not affect the quantum kinetic energy contribution, but just 
the average classical energy. In a sense, the introduction of global moves allows 
to simulate a constant-temperature classical relaxation of the average classical energy, 
superimposed on the local modifications induced by quantum fluctuations at finite temperature.

A schedule for the quantum algorithm must also be chosen. The principal drawback of the 
PIMC approach is the impossibility of simulating the quantum system directly at zero temperature. 
For this purpose, other Quantum Monte Carlo schemes should be used, like the Green Function 
Monte Carlo method \cite{GFMC_Ceperley}.
The simplest possibility for a PIMC quantum annealing schedule is, once again, to linearly adjust
the intensity of the transverse field between two given extremes $\Gamma_0$ and $\Gamma_f$,
while keeping the temperature $T$ fixed at a constant value.
More elaborated cooling protocols could be devised, but the linear strategy has 
been applied with success for the optimization of the Ising spin glass \cite{QA_ising} 
and of the Traveling Salesman Problem \cite{QA_tsp}.  
Even for this very basic schedule, four parameters, ($P,\,T, \, \Gamma_0,\,\Gamma_f$),
need to be tuned carefully in order to achieve a good performance.
A more complex field-cycling schedule will be discussed in Sec.~\ref{field_cycling:sec}.

%++++++++++++++++++++++++++++++++++++++++++++++++++++++++++++++++++++++++++++++++++++
\section{Numerical experiments: comparing classical and quantum annealing}
\label{num_exp:sec}
%++++++++++++++++++++++++++++++++++++++++++++++++++++++++++++++++++++++++++++++++++++

Let us describe now a first set of numerical experiments, performed over a single 3-SAT random 
instance with $N=10^4$ and $\alpha=4.24$, ({\it i.e.}, right at the border of the
SAT/UNSAT transition), for several choices of annealing iterations and of Trotter replicas. 
The differences among random samples extracted from the same ensemble 
become negligible for such large sizes, and we can then carry out significant experiments 
without averaging over a large set of instances. 
   
Using an efficient \textit{ad-hoc} algorithm (the WalkSAT heuristic supplemented by a
message-passing procedure \cite{Walksat,SP-Y}), 
we verified that the chosen formula (at $\alpha=4.24$) was actually satisfiable, 
as expected from theory for $\alpha<\alpha_c$.
An empirical parameter tuning was then performed for the initial temperature $T_0$. 
In the case of linear-schedule Classical Annealing (CA), and fixing the maximum number 
of iterations, we conducted several experiments with different $T_0$.
For $T_0$ smaller than the optimal value $T_0=0.3$  
the strength of the thermal fluctuations induced in the system was too small, and the dynamics 
was trapped prematurely at an early stage. On the other hand, 
for initial values larger than the optimal value, too much time was
wasted randomly wandering around the landscape at excessively high temperatures.
A similar multi-parameter optimization was performed in the case of 
linear-schedule Quantum Annealing (QA).
The best results were obtained by taking $\Gamma_0 \simeq 0.7$ and $\Gamma_f\simeq 0.001$, 
independently from the choice of the number of replicas $P$.
The optimal value of the effective quantum temperature $T_q = PT$ turned out to be 0.3, 
coincident with the optimal $T_0$ for CA, a value evidently more related to 
the statistics of the energy barriers of the formula landscape, 
than to the details of the relaxation pathway under consideration.

%------------------------------------------------------------------------------------------
\begin{figure}
\begin{centering}
\includegraphics[scale =0.42, angle=-90]{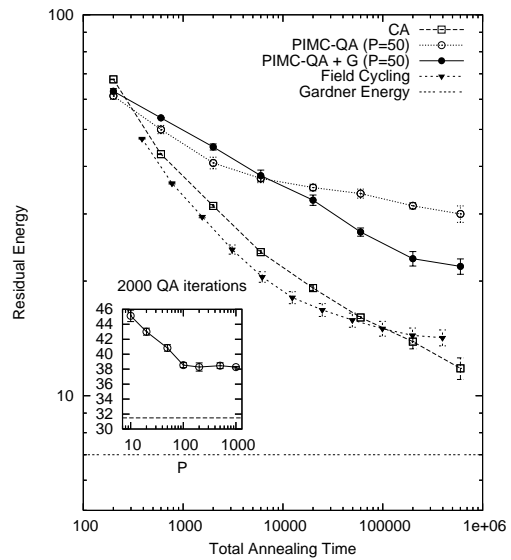}
\end{centering}
\caption{Comparison between optimal linear-schedule Classical (CA) and Quantum Annealing
(QA) for a 3-SAT problem with $N=10^4$ and $\alpha=M/N=4.24$. 
CA always performs better than QA simulated with $P=50$ Trotter replicas. 
The average performance of linear QA is worse than that of CA, even if an 
improvement in the results can be obtained by introducing global moves (G) and by increasing $P$ 
(in the inset the final average energy found by QA after 2000
iterations for increasing $P$ is plotted and compared with the average result of 
a CA of the same length, dashed line). 
The solid triangles are the data obtained by the 
field-cycling QA hybrid strategy described in Sec.~\ref{field-cycling}. }
\label{CA_vs_QA_fig}
\end{figure}
%------------------------------------------------------------------------------------------

A comparison between the performance of the optimal CA and the optimal QA at $P=50$, 
both with and without global moves, is shown in Fig.~\ref{CA_vs_QA_fig}. 
The graph shows the average residual energy (difference between the final energy reached 
by the algorithm and the konwn ground-state energy) as a function of the maximum number of 
total annealing iterations, ranging over several decades (from $10^2$ to almost $10^6$).
In each point, an average has been taken over 50 different realizations of the same experiment; 
in the case of QA, a second average was performed among the energies of the $P$ replicas, 
which are in general different. 
It can be seen that the linear-schedule CA always performs better than the linear-schedule QA, 
despite the improvement produced in the latter by the use of a large number of replicas 
(convergence is essentially reached for $P\geq 100$, see inset of Fig.~\ref{CA_vs_QA_fig}, 
but we chose $P=50$ to extend as much as possible the simulation time).
The asymptotic slope of the linear-schedule QA curves seems indeed to be definitely less steep 
than that of CA, independently of the number of replicas involved in the simulation and of
the use of global moves. 
This behavior is strikingly worse than that observed in both the Ising spin glass 
and the TSP cases \cite{QA_ising,QA_tsp}. 
A detailed study of the influence of landscape geometry might be helpful in 
order to achieve a better understanding of this different performance success. 
It should be remarked that large differences exist between the organization of the 
optimal states in the combinatorial problems analyzed so far. 
For instance, it is generally believed that the ground state properties of the 
TSP \cite{TSP_theory} are well described by a replica-symmetric Ansatz 
(all the ground state assignments are in a single broad valley), while 
replica-symmetry-breaking (\textit{i.e.}, clustering of ground states) is needed to achieve a good
characterization of the 3-SAT zero-temperature properties \cite{SP_long}.
Furthermore, the states visited by both CA and QA lie well above the 
glassy threshold (Gardner energy) for the 3-SAT instance, and very few details are known 
about the nature of such highly excited states.
An attempt at characterizing the local topology of the visited configurations will be 
made in Sec.~\ref{autocorr:sec}.

%++++++++++++++++++++++++++++++++++++++++++++++++++++++++++++++++++++++++++++++++++++
\section{Monte Carlo time dynamics and evolutionary analogy}
\label{MC_dynamics:sec}
%++++++++++++++++++++++++++++++++++++++++++++++++++++++++++++++++++++++++++++++++++++

\begin{figure}
\includegraphics[scale =0.42]{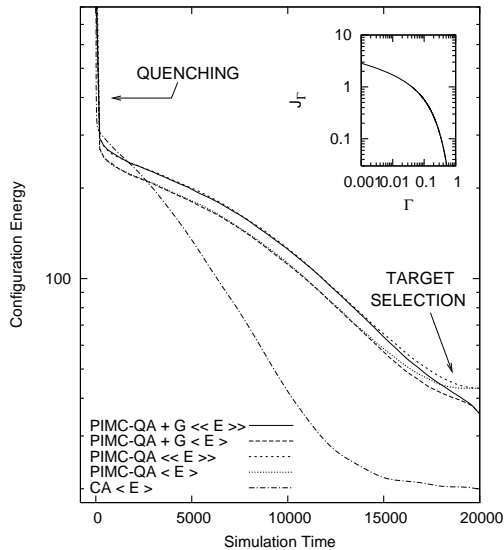}
\caption{Energy evolution during Quantum Annealing, compared to Simulated Annealing. 
The variation of the averages $\left<E\right>$ (average best replica)
and $\left<\left<E\right>\right>$ (average of the average replica)
is shown as a function of the simulation time, for a set of experiments with $P=50$ and
a 2000 annealing iterations. 
The inset shows the time-dependent value of the coupling $J_\Gamma$.
Three different regimes can be distinguished, which will be called quenching, search (driven by 
quantum fluctuations) and target selection.}
\label{Energy_Time_fig}
\end{figure}

Let us now analyze in more detail the time-evolution of the energy during the annealing dynamics. 
We denote by $\left<\left<E\right>\right>$ the configuration energy averaged over different 
experiments \textit{and} Trotter replicas (this is the energy reported everywhere in 
Fig.~\ref{CA_vs_QA_fig}); the average among different experiments of the \textit{best} replica 
energy will be, on the other hand, denoted as $\left<E\right>$. 
In Fig.~\ref{Energy_Time_fig}, the Monte Carlo ``time'' evolution profiles of 
$\left<\left<E\right>\right>$ and $\left<E\right>$ are shown for a linear-schedule QA 
(2000 iterations long).
%
%------------------------------------------------------------------------------------------
%------------------------------------------------------------------------------------------

The strength of the transverse field, and hence of the quantum coupling $J_\Gamma$ given by 
Eq.~(\ref{transverse_coupling}) (see inset of Fig.~\ref{Energy_Time_fig}),
determines the relative importance of the classical and quantum terms in the Hamiltonians 
(\ref{quantum_ham}) and (\ref{suzuki_ham}), and its variation determines the transition 
between the three following observed regimes:
\begin{itemize}
\item [$1$.] (\textit{Quenching phase}, $\Gamma\approx \Gamma_0\simeq 0.7$, $J_{\Gamma}\simeq 0$). 
The quantum system described by the Hamiltonian (\ref{quantum_ham}) is quenched 
at temperature $T_q$ in presence of a strong external transverse field. 
The system enters an incoherent mixture of states. 
From the point of view of the PIMC simulation of Hamiltonian (\ref{suzuki_ham}), 
when the coupling $J_\Gamma$ is small, each replica behaves as if roughly independent from 
the others. 
The algorithm is then effectively simulating $P$ parallel Monte Carlo dynamic processes 
at a constant temperature $T_q$. Both the local and global moves acceptances are extremely high in
this transient regime.
The replicas assume different configurations at similar energies, compatible with the 
quenching level at temperature $T_q$. 
\item [$2$.] (\textit{Search phase}).
After the abrupt out-of-equilibrium quenching phase, quantum features like delocalization, 
interference and tunneling should help the system to evolve towards attractive low-energy 
configurations. 
Looking at the Suzuki-Trotter Hamiltonian in Eq.\ (\ref{suzuki_ham}), with increasing coupling 
strength, the fluctuations of the different replicas become correlated. 
Some spin flips that would have been unlikely in absence of the 
replica-interaction term can now be accepted, thanks to the kinetic term, and several replicas 
can enter in configurations generally not visited by typical CA trajectories. 
The acceptance ratio is constantly higher than in the CA case and around the 25\%, when 
considering local moves. It is negligible, on the other hand, for global moves.
\item[$3$] (\textit{Target selection phase}, $\Gamma\approx \Gamma_f$, $J_{\Gamma}$ large).
When the transverse field vanishes, quantum fluctuations are gradually switched off. 
The system selects then a low-energy target state and collapses completely into it. 
In this classical regime, when the transverse coupling $J_\Gamma$ becomes very strong, 
only local spin flips tending to align the largest number of replicas can be accepted. 
All the replicas converge then towards the same configuration, corresponding to the one visited 
that has minimum energy, if the parameters have been carefully tuned. 
The acceptance for local moves falls essentially to zero, but, if global moves
are allowed, a further energy reduction of moderate entity can be observed thanks to 
small-range classical oscillations (the global moves acceptance reaches now approximately 
the 20\% level).
\end{itemize}
Considering this three-piece scenario (that will be confirmed by the autocorrelation 
and geometrical analysis of Sec.~\ref{autocorr:sec}), 
the simulated QA could be described as a very basic 
kind of \textit{evolutionary search} \cite{GenAl_book,Holland_classic}.
The $P$ replicas can be seen as a population of individuals, the spin configuration of each 
replica as its genotype, and the classical Hamiltonian (\ref{SAT_ham}) as a fitness function. 
The simulation is not simply equivalent to the selection of the best among $P$ 
``restarts'' \cite{MontanariZecchina}. because here contiguous replicas can ``mate'', 
exchanging sequences of their genotype thanks to the duplicating action of the transverse coupling.
A global decrease of the average fitness is induced by the proliferation of spin patterns 
typical of low-energy replicas (the so-called Holland \textit{schemata} \cite{Holland_schemata}).
Suppose now that some new exceptional individual appears when $J_\Gamma$ is already 
considerably large. The chance of survival of the spin pattern responsible of its 
superior fitness  would be rather small, because it would be overwritten with high probability 
by the corresponding subsequences in the most widespread configuration. 
The population tends then at a certain point to collapse toward a group of ``identical twins''. 
Global moves cannot alter significantly this picture, because they do not
cure the problem of the lack of genetic diversity.
In standard evolutionary search, the available gene pool is constantly renewed by crossover 
techniques, but their introduction would not be physically justifiable in our PIMC scheme. 
However, one is in principle allowed to increase the mutation rate by switching on 
again the quantum fluctuations, a strategy that will be investigated in 
Sec.~\ref{field_cycling:sec}.

%++++++++++++++++++++++++++++++++++++++++++++++++++++++++++++++++++++++++++++++++++++++++
\section{Autocorrelation analysis and landscape probing} 
\label{autocorr:sec}
%++++++++++++++++++++++++++++++++++++++++++++++++++++++++++++++++++++++++++++++++++++++++

A better characterization of the dynamical behavior can be obtained by looking at the 
correlations among spin configurations at different times, and by probing the geometry 
of the neighborhood of the visited configurations.

\begin{figure}
\begin{centering}
\includegraphics[scale =0.42, angle=-90]{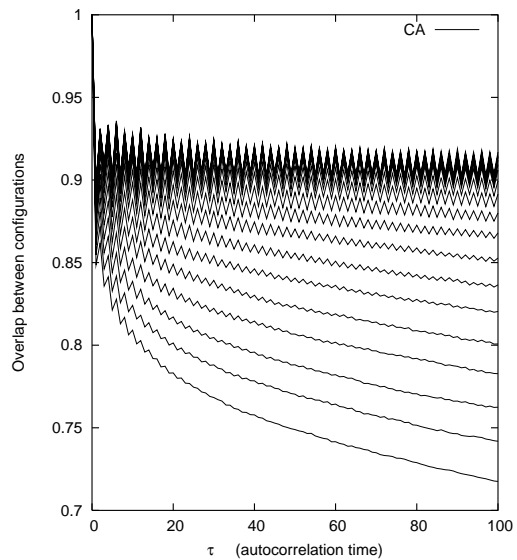}
\end{centering}
\caption{Autocorrelation function $K(t^*,\tau)$ for CA. 
The different curves represent the decay with $\tau$ of several fixed-simulation-time 
snapshots of the autocorrelation function, for a CA experiment; 
$t^*$ is varying in the plot at fixed intervals between 200 and 2000, from bottom to top.}
\label{Tau_against_overlap_C}
\end{figure}

\begin{figure}
\begin{centering}
\includegraphics[scale =0.42, angle=-90]{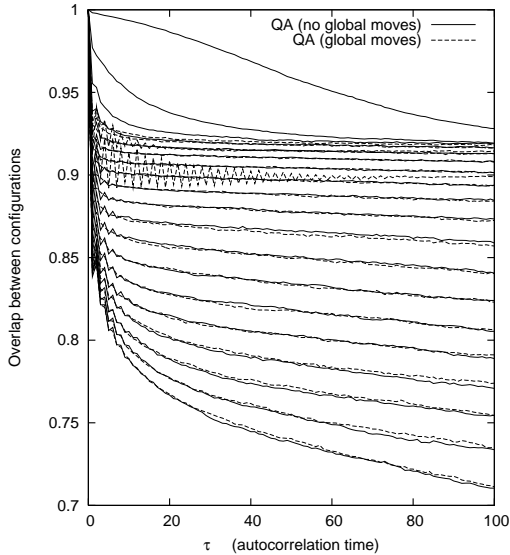}
\end{centering}
\caption{Autocorrelation function $K(t^*,\tau)$ for QA. 
The different curves represent the decay with $\tau$ of several fixed-simulation-time 
snapshots of the autocorrelation function, for QA experiments with and without global moves; 
$t^*$ is varying at fixed intervals between 200 and 2000, from bottom to top. 
The target selection phase is characterized by a complete collapse into a single configuration, 
while damped classical oscillations around the equilibrium position are allowed if global 
moves are introduced.}
\label{Tau_against_overlap_Q}
\end{figure}

Let us denote by $\{S_i(t)\}$ the instantaneous spin configuration of the sample 
3-SAT-formula at time $t$.
An \textit{autocorrelation function} $K(t,\tau)$ can be defined as:
\begin{equation}\label{autocorrelation}
K(t,\tau) = \left\langle\frac{1}{N}\sum_{i=1}^N S_i(t)S_i(t-\tau)\right\rangle \;,
\end{equation}
where an average over different realizations of the dynamics (and over replicas in the QA case)
has to be understood.
The autocorrelation function $K(t,\tau)$ allows us to visualize in a compact way the typical 
behavior of the overlap between two spin assignments at different evolution instants. 
In Figs.~\ref{Tau_against_overlap_C} and \ref{Tau_against_overlap_Q} 
we plot $K(t,\tau)$ as a function of the autocorrelation time $\tau$ for several fixed 
values $t^*$ of the simulation time $t$, for the CA and QA dynamics, respectively
(in the plots, $t^*$ is generally increasing from bottom to top, see later
in this section).
The results shown are averaged over 500 different runs, each of 2000 annealing iterations
(and over $P=50$ replicas, in the case of QA).
A $K(t^*,\tau)$ which decays fast with $\tau$ indicates that at time $t^*$ the configuration 
is still rapidly evolving, and that at every time-step a large number of spins is being flipped; 
when the local stability is reached, on the other hand, $K(t^*,\tau)$ assumes a flat 
(or periodic) profile, indicating that the system has entered into some attracting configuration 
(or limit cycle).

%------------------------------------------------------------------------------------------
%------------------------------------------------------------------------------------------
%------------------------------------------------------------------------------------------

Looking at Fig.~\ref{Tau_against_overlap_C}, for CA, the self-overlap between $\{S_i(t^*)\}$ 
and $\{S_i(t^*-\tau)\}$ grows with $t^*$. 
With increasing $t^*$, a periodic behavior of $K(t^*,\tau)$ begins to appear over initially 
short but gradually increasing decay times $\tau$. In the final part of the classical 
relaxation, about 20\% of the variables are still allowed to flip at each iteration, even if 
the average energy is no longer changing. 
The strongly regular oscillations of $K(t^*,\tau)$, as well as the non-vanishing asymptotic 
spin flip acceptance ratio, suggest that the system gets trapped into a very small portion of 
the phase space, and that a fraction of the variables is still allowed to 
fluctuate, but only cyclically repeating a limited amount of sequences of flips. 

\begin{figure}
{\centering
\vspace{2ex}
\includegraphics[scale =0.42, angle=-90]{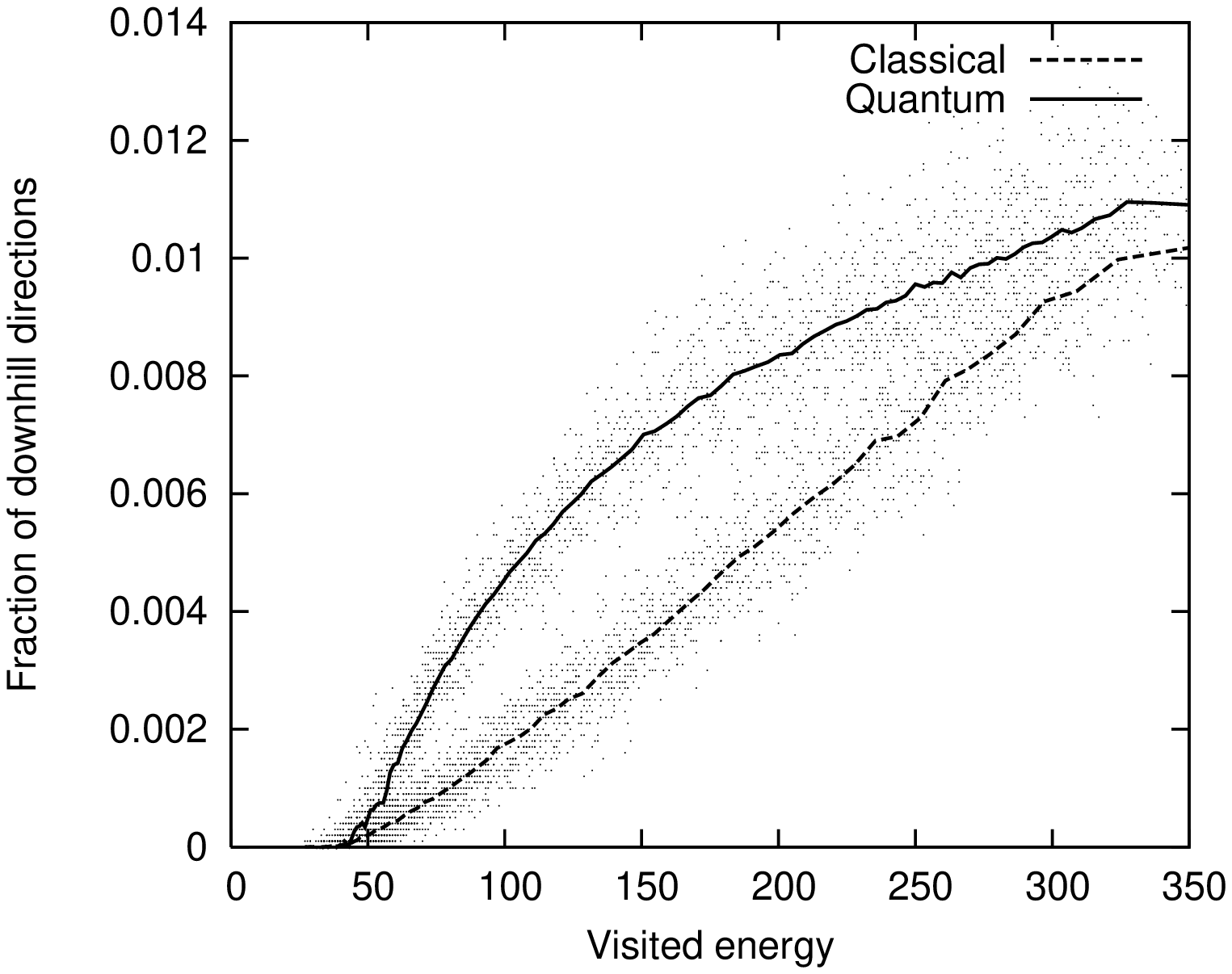}
\includegraphics[scale =0.42,angle=-90]{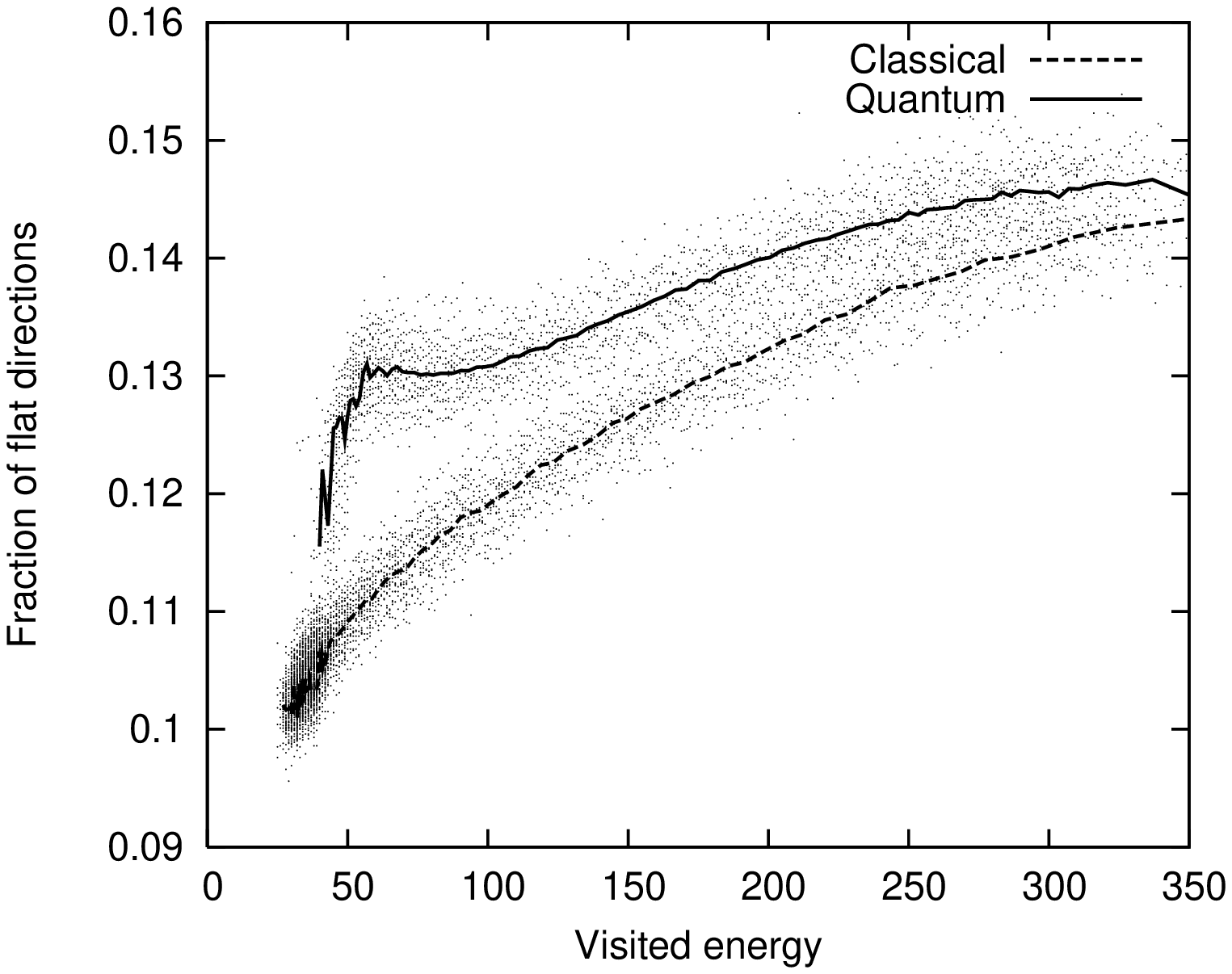}\\
\includegraphics[scale =0.42, angle=-90]{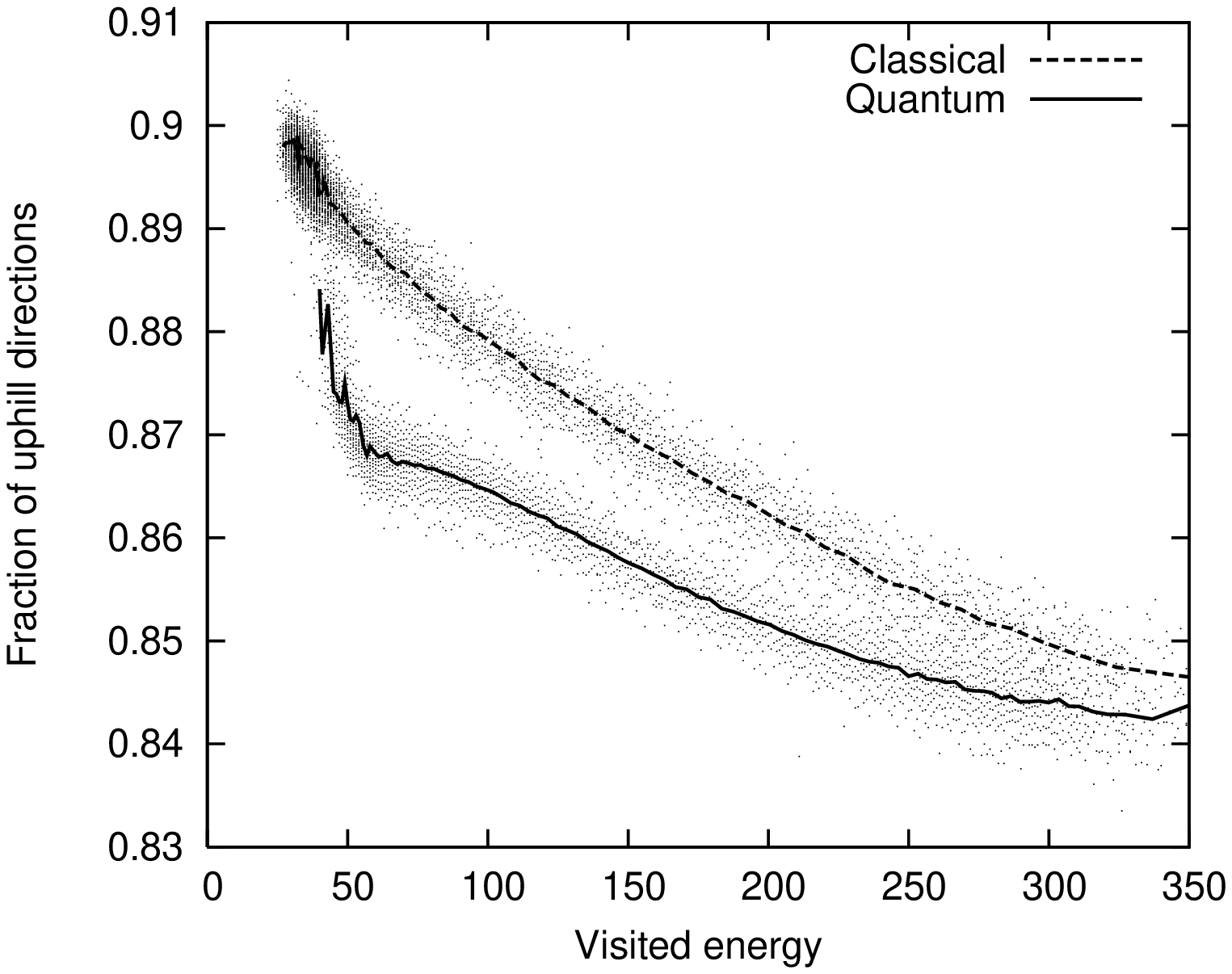}
}
\caption{The local geometry of the visited regions of the phase space is probed by counting
the fraction of directions in which the energy variation is negative, null or positive. Although
both CA and QA get trapped in a local minimum, the quantum evolution tends to visit ``valleys'' 
that, at the same energies than CA, are more flat and with a larger number of downhill directions.}
\label{landscape_probing_fig}
\end{figure}

It is possible to characterize the states that are actually blocking the system, 
by looking at the neighborhood geometry during the dynamics.
When sitting in a configuration at a given energy $E$, one studies the energy change
$\Delta E$ upon flipping all the $N$ spins: $\Delta E$ can be positive (uphill direction), 
vanishing (flat direction) or negative (downhill direction).  
The data relative to the configurations sampled by a large number of linear CA runs
(each of 2000 annealing iterations, and always over the same $N=10000$ sample) 
are shown in Fig.~\ref{landscape_probing_fig}, where the fractions of downhill, flat and uphill 
directions are plotted (from top to bottom) against the energy of the visited configurations.
One sees that the number of downhill directions falls to zero when the lowest energies are 
approached, indicating that the CA dynamics ends in a local minimum. 
The number of remaining flat directions is compatible with the observed amplitude of the 
oscillations shown in Fig.~\ref{Tau_against_overlap_C} for large $t^*$. 

Moving now to the analysis of QA, it must be remarked that during most of the simulation time 
the self-overlap among configurations at different times is considerably smaller than 
in the case of CA, and the spin-flip acceptance ratio is larger. 
Both these properties hints at a regime characterized by a rapid and disordered evolution 
of all the replica spin-configurations. 
The self-overlap increase becomes faster upon reducing the transverse magnetic field, 
because the pseudo-evolutionary replication of the ``good'' spin patterns, operated 
by the coupling (\ref{transverse_coupling}), has also a stabilizing effect on them. 
If only local moves are performed (solid curves in Fig.~\ref{Tau_against_overlap_Q}), 
no trace of asymptotic periodic behavior is found, and all the replica configurations 
reach continuously a full overlap with a single final configuration.
% 
%------------------------------------------------------------------------------------------
%------------------------------------------------------------------------------------------
%
When global moves are switched on, at large $t^*$, after reaching a plateau of approximately 
0.93, the target selection phase starts, but now the self-overlap reduces slightly and 
damped classical-like oscillation set in. 
Looking at Fig.~\ref{landscape_probing_fig}, it is possible to see that the states reached
at the end of the QA schedule are very close to local minima, with an extremely small 
number of downhill directions. 
Once again, we assist to a trapping phenomenon, and the global moves can only produce 
a local optimization around the final configuration reached when just local moves are used.
Even if the diverging transverse coupling (and the consequent vanishing of the local acceptance) 
modifies in a noticeable way the behavior of the autocorrelation function, 
the blocking mechanism acting in the target selection phase of QA is then perfectly analogous
to that of CA.

Nevertheless, the analysis of the local geometry, shown in Fig.~\ref{landscape_probing_fig},
highlights important differences between the CA and the QA dynamics.
The phase space region explored by the two algorithms are quite different. 
At the same value of the energy, the quantum system is visiting configurations with
a significantly larger fraction of downhill and flat directions. 
If an abuse of language is tolerated, one could say that the CA follows narrow canyons, 
where the number of directions bringing to a decrease in energy is limited, while the 
QA prefers to explore the edges of mid-altitude plateaus.
This phenomenon, which seems to be a genuinely quantum feature captured by the PIMC simulation, 
is strongly reminiscent of what happens in continuous space, where the choice of
broader potential wells allow the system to reduce the kinetic contribution to the total energy 
(curvature-induced effects are also well known in the theory of instanton 
tunneling \cite{kleinert}).
Quite interestingly, the typical number of spins which are different among the various Trotter 
replicas is of the order of the number of flat directions. This means that all the configurations 
simultaneously taken by the quantum system belong to a single broad landscape valley, which 
is explored in all its wideness by the quantum system. 
When the transverse field intensity is vanishing, the target selection transition toward 
a classical regime takes place, and the number of uphill directions increases abruptly, 
indicating that the dynamical collapse is paralleled by a change in the local 
landscape topology. 
The poor performance of QA in the present 3-SAT case could be then explained by the existence 
of broad basins of attraction strewn with deceptive and highly attractive sinks, that, 
unlike the cleavages preferred from the very beginning by the CA, prevent access to 
lower energy sectors. To conclude, we remark that genetic-like search is also known to be 
strongly perturbed by the presence of multiple and scattered wells \cite{dejong}.

%++++++++++++++++++++++++++++++++++++++++++++++++++++++++++++++++++++++++++++++++++++++++
\section{Field-cycling strategies}\label{field-cycling}
\label{field_cycling:sec}
%++++++++++++++++++++++++++++++++++++++++++++++++++++++++++++++++++++++++++++++++++++++++
\begin{figure}
{\centering
\vspace{2ex}
\includegraphics[scale =0.42,angle=-90]{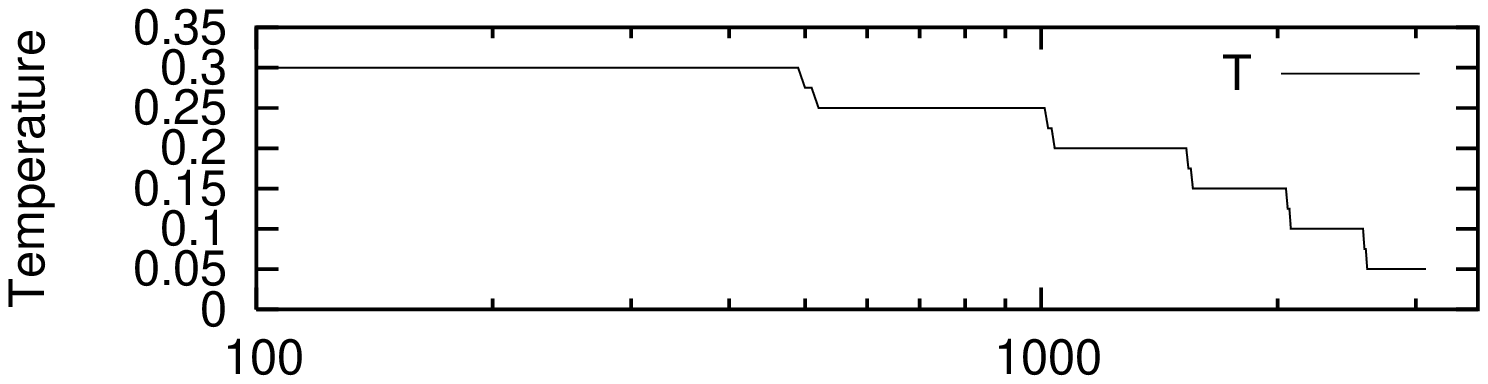}\\
\includegraphics[scale =0.42,angle=-90]{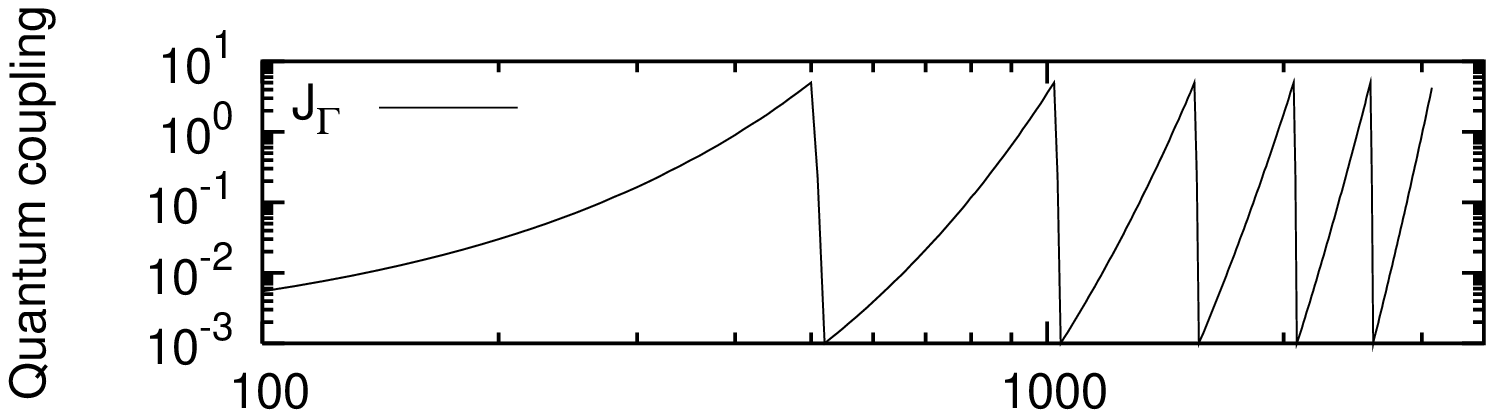}\\
\includegraphics[scale =0.42]{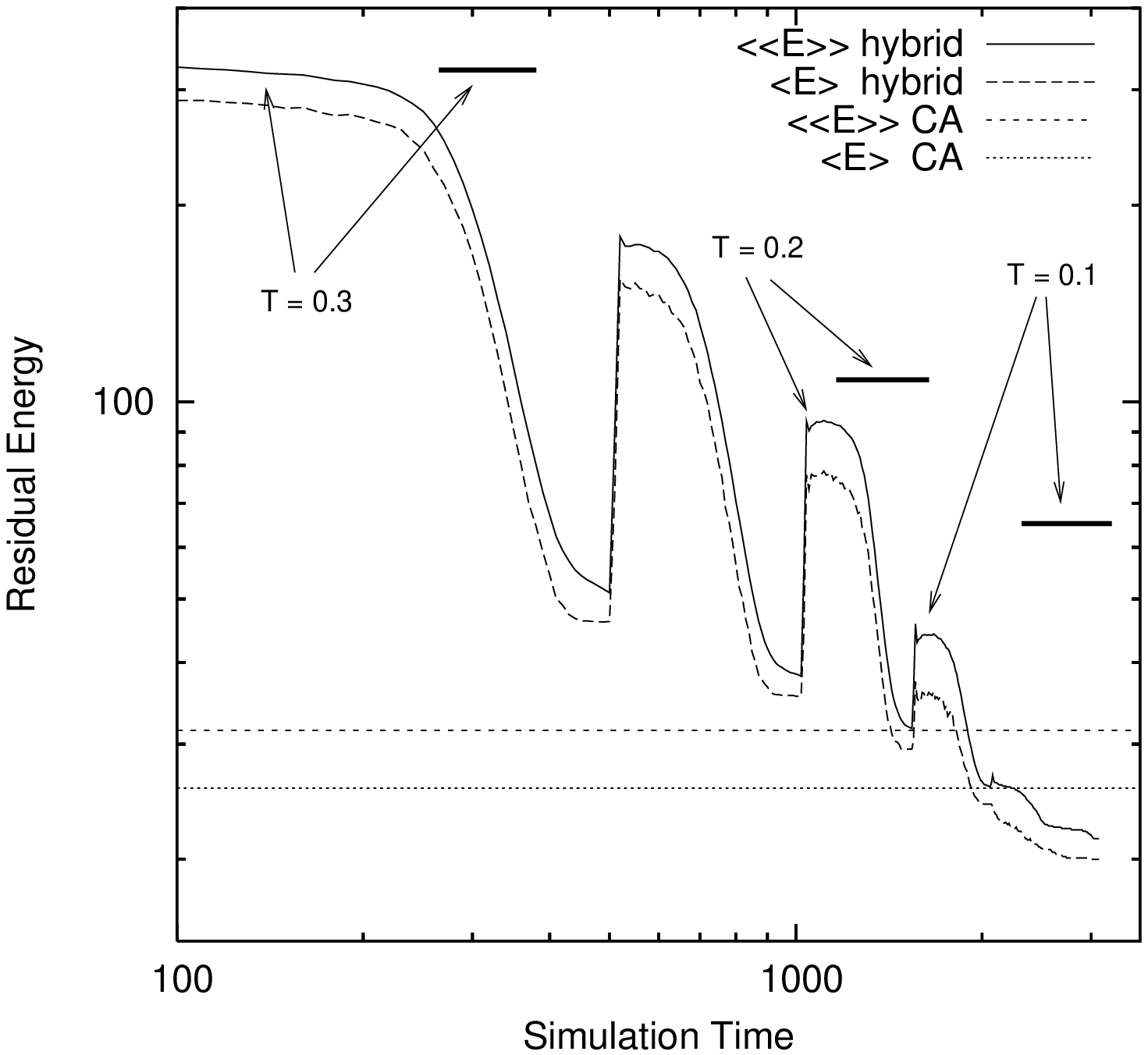}
}
\caption{Energy evolution during a field-cycling hybrid strategy.
The strength of the transverse coupling $J_\Gamma$
is varied cyclically between the values 0.001 and 5, by adjusting the value of the magnetic field.
The effective temperature $T_q$ is kept constant during each field ramp, but is reduced in a 
stepwise way among different ramps, from the initial value of 0.3 down to 0.05.
Each ascending field-ramp unfreezes the system from a previously reached target state,
and after a short transient regime, a new search phase is entered. 
The starting plateaus have energy values increasingly smaller than the
quenching level at the new simulation temperature (the arrows in the
graphs indicates the quenching level and the hybrid strategy plateau
at a given value of the temperature). Each new target
state has a better energy than the preceding one, and the final
average energy is better than the value reachable by large classical fluctuations.}
\label{field_cycling_fig}
\end{figure}
%------------------------------------------------------------------------------------------

After the final target selection, a new quantum search phase can be started by 
switching on again the field $\Gamma$ and restoring then a regime dominated 
by quantum fluctuations.

In a first possible experimental setup (results not shown), after a linear descending 
ramp from $\Gamma_0\simeq 0.7$ to $\Gamma_f\simeq 0.001$, the field is raised 
smoothly to the initial value $\Gamma_0$ and then back again to $\Gamma_f$,
while keeping the effective quantum temperature $T_q=0.3$ constant. 
Many such field-cycles can be chained one after the other. 
It turns out that the linear-schedule QA dynamics is perfectly time-reversible, 
and that all the ascending and descending ramps produce almost exactly the same 
time-evolutions for the energy 
(this phenomenon is somehow analogous to the well known memory and rejuvenation
effects in glassy out-of-equilibrium dynamics \cite{bouchaud}). 

Even if the restoration of quantum fluctuations 
allows the system to escape from the target local minimum, 
a further ingredient is needed in order to achieve better results. 
A possible expedient to avoid a complete re-initialization 
is to slightly reduce the temperature $T$ before each new field-ramp; 
the temperature is still kept constant during each individual ramp, realizing thus a 
hybrid strategy (a linear-schedule CA, superposed with linear-schedule QA cycles over 
a shorter time-scale).
The time-dependence of the energy along a successful field-cycling 
scheme with a total length of 3100 MC iterations is presented in Fig.~\ref{field_cycling_fig}. 
The extrema $\left(\Gamma_0^{(i)}, \Gamma_f^{(i)}\right)$ of the 
$i$-th field-ramp are selected in order to have a smooth variation 
of $J_\Gamma$, when the temperature changes discontinuously among two different ramps. 
The coupling is then varied regularly and cyclically between the values 
$J_{\Gamma, 0}\simeq 0.001$ (corresponding to $\Gamma_0\simeq 0.7$ when $T_q=0.3$) 
and $J_{\Gamma, f}\simeq 5$ (corresponding to $\Gamma_0\simeq 0.001$ when $T_q=0.3$). 
The temperature $T_q$, finally, is reduced from 0.3 to 0.05 as shown in 
the upper panel of Fig.~\ref{field_cycling_fig}.
Unlike the case of constant temperature field-cycling, 
after each ascending ramp and a short transient phase, we reach now new plateaus, 
lying at a distance progressively larger from the
energy level that would be obtained by performing a classical quenching 
at the same temperature (see the arrows in Fig.~\ref{field_cycling_fig}). 
Lower-lying target states are then selected after each descending ramp. 
Over short time scales (number of MC iterations approximately smaller than 
200000, when taking $P=50$), such a hybrid field-cycling strategy performs definitely better
than a purely classical one. The same experiment has been repeated 
in absence of the transverse magnetic field, and with the same number of 
(now completely decoupled) Trotter replicas. 
The average over the runs and over the Trotter replicas 
$\left<\left<E\right>\right>_{\mbox{\tiny{CA}}}$,
and the average over the runs of the best replica energy 
$\left<E\right>_{\mbox{\tiny{CA}}}$ have been computed.
It is possible to see from Fig.~\ref{field_cycling_fig} that 
$\left<\left<E\right>\right>_{\mbox{\tiny{hybrid}}}$ obtained with the hybrid strategy 
lies clearly below $\left<E\right>_{\mbox{\tiny{CA}}}$, indicating that quantum 
effects give access to states that can hardly be reached in the same time even by 
rare large classical fluctuations. 
Quantum restarts are then more effective than classical restarts \cite{MontanariZecchina}, at 
least when short schedules are taken in account.

The analysis of purely classical cycling experiments confirms the importance of 
quantum fluctuations.
The fact that 
$\left<E\right>_{\mbox{\tiny{CA}}} > \left<\left<E\right>\right>_{\mbox{\tiny{hybrid}}}$ 
shows already that the superior performance is not simply due to the gradual reduction 
of the initial temperature among the different ramps. 
Furthermore, the concatenation of short classical annealings,
each new one starting from a lower $T_0$ and going down to $T_f = 0$, provide 
a worse final energy than a single classical annealing of the same length. 
Indeed, even if the $i$-th descending classical annealing ramp (starting from $T_0^{(i)}$) 
produces a configuration with an energy lower than in the random case,
the $i$-th fast ascending ramp to the temperature $T_0^{(i+1)}<T_0^{(i)}$ of the 
$(i+1)$-th descending ramp completely erase now any positive effect. 
The system is indeed pushed at the energy level typical of a fast quenching 
at the temperature $T_0^{(i+1)}$ and, unlike the quantum case, 
the memory of the previously visited low-energy configurations is completely lost.
The presence of the kinetic energy term is therefore crucial for the survival of  
``seed patterns'', peculiar of the target configurations obtained at the end of the 
$i$-th descending ramp, from which new and better low-lying configurations can be grown 
during the following $(i+1)$-th stage.
In the genetic algorithm jargon, one could say that the ascending ramps renew 
the available gene pool without destroying completely the highly fit genes 
of the preceding-step generation.
 
The situation is however different for larger time scales 
(MC iterations larger than 200000 for $P=50$). Longer field-cycling schedules are
obtained by simply rescaling with a constant factor the duration of all
the ramps in a shorter schedule. The asymptotic slope of the field-cycling cooling curve in 
Fig.~\ref{CA_vs_QA_fig} becomes remarkably similar to the other QA cases. 
A local topology analysis analogous to the one of the previous section shows indeed that
the valleys explored by the field-cycling strategy are flat and open as the ones found 
in the simpler linear QA case.  
If the reduction of temperature allows the system to explore the landscape 
at different length scales \cite{bouchaud} and to find then better target configurations, 
the attractive power of the visited high-lying local minima continues to be very strong, and 
lower energy regions remain fundamentally inaccessible. 

%++++++++++++++++++++++++++++++++++++++++++++++++++++++++++++++++++++++++++++++++++++++++
\section{Discussion and conclusions}
\label{conclusions:sec}
%++++++++++++++++++++++++++++++++++++++++++++++++++++++++++++++++++++++++++++++++++++++++

We have conducted a detailed comparison of the performance of Path-Integral Monte Carlo
(PIMC) Quantum Annealing (QA) strategies, against a standard thermal simulated 
annealing (CA), on a very hard instance of a classical NP-complete problem, 3-SAT.
Contrary to the successes previously obtained by the same technique on 
Ising spin glasses \cite{QA_ising} and on an instance of the Traveling Salesman Problem 
\cite{QA_tsp}, QA performs here definitely {\em worse} than CA: the ultimate
large simulation time behavior of QA shows a poorer slope against inverse 
annealing rate than CA (see Fig.~1).

In the course of this instructive negative example of QA performance (perhaps even 
more instructive because it is negative), we gained some experience and
insight on the peculiar dynamical relaxation process behind the PIMC-QA algorithm. 
In particular, we saw, from overlap autocorrelation analysis, that the quantum
algorithm leads to ``selection'', in the course of annealing, of a target configuration
over which all the different replicas eventually collapse, as the transverse coupling 
$J_\Gamma$ induced by the quantum term $-\Gamma\sum_i\sigma_i^x$ grows to infinity.
As a byproduct, we also realized that restarting repeatedly quantum fluctuations using such
target configurations as intermediate steps leads to a hybrid strategy that definitely
improves over the bare linear schedule QA, with results comparable to CA for short
simulation times. Nevertheless, even with these considerable improvements of QA,
the slope of the annealing curve seems to be essentially unaffected by refinements, 
and definitely worse than that of CA. This apparent ``intrinsic'' nature of the 
worse performance of PIMC-QA over CA suggests that the 3-SAT landscape (at least, 
when single spin-flip moves are considered) is in some way more difficult for the quantum
algorithm than for the classical one, presumably due to the
fact that the flat and open landscape sectors that are peculiarly selected by 
the quantum relaxation bring in the 3-SAT case toward ``dead ends'', hiding dangerous sink traps.

Finally, we should stress that both PIMC-QA and CA are definitely worse than ad-hoc 
local search algorithms, notably WALKSAT \cite{Walksat}. 
One probably important feature that is missing to both
CA and PIMC-QA, but known to be important in WALKSAT and similar local heuristics, is
the so-called {\em focusing}. That is the fact that the proposed random flips should be
restricted to spins that are exclusively involved in currently UNSAT clauses 
(according to the rule of thumb ``if it is not broken, don't fix it''). 
On the contrary, standard Metropolis sampling, as implemented in both PIMC-QA and CA, 
does not discriminate among the candidate spin flips, and only considers the 
corresponding energy change $\Delta E$ (see Eq.~\ref{acceptance}). 
This suggests that a possible better way of implementing QA is through a 
Green's Function Monte Carlo (GFMC) algorithm, whereby a suitable kinetic term and appropriate 
guiding functions, via importance sampling, can in principle take care of some form of focusing.
Another important advantage of GFMC over PIMC would be the fact that annealing
could be performed strictly at $T=0$, whereas the unavoidably finite temperature $T$ has clearly
an influence on the PIMC-QA dynamics (see Fig.~\ref{field_cycling_fig}).

In conclusion, while showing that statements such as ``quantum is better'' have not 
necessarily anything to do with physical reality -- at least for arbitrary choices of the 
problem landscape and of the associated elementary moves --, 
our work still leaves many important questions open. 
One should strive for a reliable {\em predictive} theory that, 
taking as inputs the appropriate features of the complex energy landscape for the problem 
at hand, or for a class of problems, would be able to anticipate if, how, and where 
``quantum is better''. 
Closely connected with this is the ability in designing quantum kinetic 
Hamiltonian terms that implement \textit{ad hoc} local moves particularly efficient in
exploring a given geometry. 

We thank Michal Kol\'a\v{r}, Mario Rasetti, Lorenzo Stella, Osvaldo Zagordi and Riccardo 
Zecchina for stimulating discussions.

%%%%%%%%%%%%%%%%%%%%%%%%%%%%%%%%%%%%%%%%%%%%%%%%%%%%%%%%%%%%%%%%%%%%%%%%%%%%%%%%%%%%%%%%%%%%%
% Bibliography
%%%%%%%%%%%%%%%%%%%%%%%%%%%%%%%%%%%%%%%%%%%%%%%%%%%%%%%%%%%%%%%%%%%%%%%%%%%%%%%%%%%%%%%%%%%%%

\end{document}